\documentclass[rnote]{aa} 

\usepackage{txfonts}
\usepackage{graphicx}
\usepackage{natbib}

\bibpunct{(}{)}{;}{a}{}{,}

\newcommand{\ekuma}{EK\,UMa}

\newcommand{\rosat}{\textit{ROSAT}}
\newcommand{\einstein}{\textit{Einstein}}
\newcommand{\euve}{\textit{EUVE}}

\begin{document}

\title{A long-term optical and X-ray ephemeris of the polar EK~UMa} 

\author {K.\ Beuermann \inst{1}, J.\ Diese \inst{2}, S.\ Paik
  \inst{3}, A.\ Ploch \inst{2}, J.\ Zachmann \inst{2}, A.~D.\ Schwope
  \inst{4} \& F.~V.\ Hessman \inst{1} }


\institute{Institut f\"ur Astrophysik, G\"ottingen, Germany, beuermann@astro.physik.uni-goettingen.de, hessman@astro.physik.uni-goettingen.de
\and
Max-Planck-Gymnasium, G\"ottingen, Germany, diese@gmx.de, icarebooten@web.de, jzachmann@t-online.de
\and
Felix-Klein-Gymnasium, G\"ottingen, Germany, swishfugu00@yahoo.co.kr
\and
Astrophysikalisches Institut Potsdam, Potsdam, Germany, aschwope@aip.de}

\date{Received 26 June 2009 / accepted 20 August 2009}
 
\authorrunning{K. Beuermann et al.}  \titlerunning{Long-term ephemeris of EK~UMa}

\abstract {}
{ 
We searched for long-term period changes in the polar \ekuma\ using
new optical data and archival X-ray/EUV data.}
{ 
An optical ephemeris was derived from data taken remotely with the
MONET/N telescope and compared with the X-ray ephemeris based on
\einstein, \rosat, and \euve\ data. A three-parameter fit to the
combined data sets yields the epoch, the period, and the phase offset
between the optical minima and the X-ray absorption dips.  An added
quadratic term is insignificant and sets a limit to the  period
change.}
{ 
The derived linear ephemeris is valid over 30 years and the common
optical and X-ray period is $P=0.0795440225(24)$\,days. There is no
evidence of long-term $O-C$ variations or a period change over the
past 17 years ($\Delta P = -0.14\pm 0.50$\,ms). We suggest that the
observed period is the orbital period and that the system is tightly
synchronized. The limit on $\Delta P$ and the phase constancy of the
bright part of the light curve indicate that $O-C$ variations of the
type seen in the polars DP\,Leo and HU\,Aqr or the pre-CV NN\,Ser do
not seem to occur in \ekuma. The X-ray dips lag the optical minima by
$9.5^\circ\pm0.7^\circ$ in azimuth, providing some insight into the accretion
geometry.}
{}

\maketitle

\section{Introduction}

Long-term studies of cataclysmic variables and related objects have
shown that variations in the orbital period occur in some of them,
which are not understood or badly so
\citep[e.g.][]{beuermannpakull84,pandeletal02,schwopeetal02,brinkworthetal06,schwarzetal09}.
We set out to search for period changes in other eclipsing and
non-eclipsing CVs using remote observations with the MONET/N telescope
as part of a secondary school research project.

The polar EK~UMa is the 19 mag optical counterpart of the bright {\it
  EINSTEIN\/} X-ray source 1E1048.5+5421 discovered by
\citet{morrisetal87}. No optical photometry besides that in the
discovery paper is in the literature, but an accurate ephemeris based
on \rosat\ X-ray data was derived by \citet{schwopeetal95}. We present
new optical photometry and supplement the \rosat\ X-ray timings by an
analysis of the \euve\ deep survey DS/S light curves taken in 1994,
1998, and 1999. The combined data allow us to establish a common
long-term ephemeris that extends 30 years back to the \einstein\ era
and accounts for a systematic phase offset between the X-ray
absorption dips and the optical cyclotron minima. No period change was
found, but we derive a tight limit on $O-C$ variations and associated
period changes over the past 17 years.

\section{Optical observations}

\begin{figure}[t]
\includegraphics[width=8.7cm]{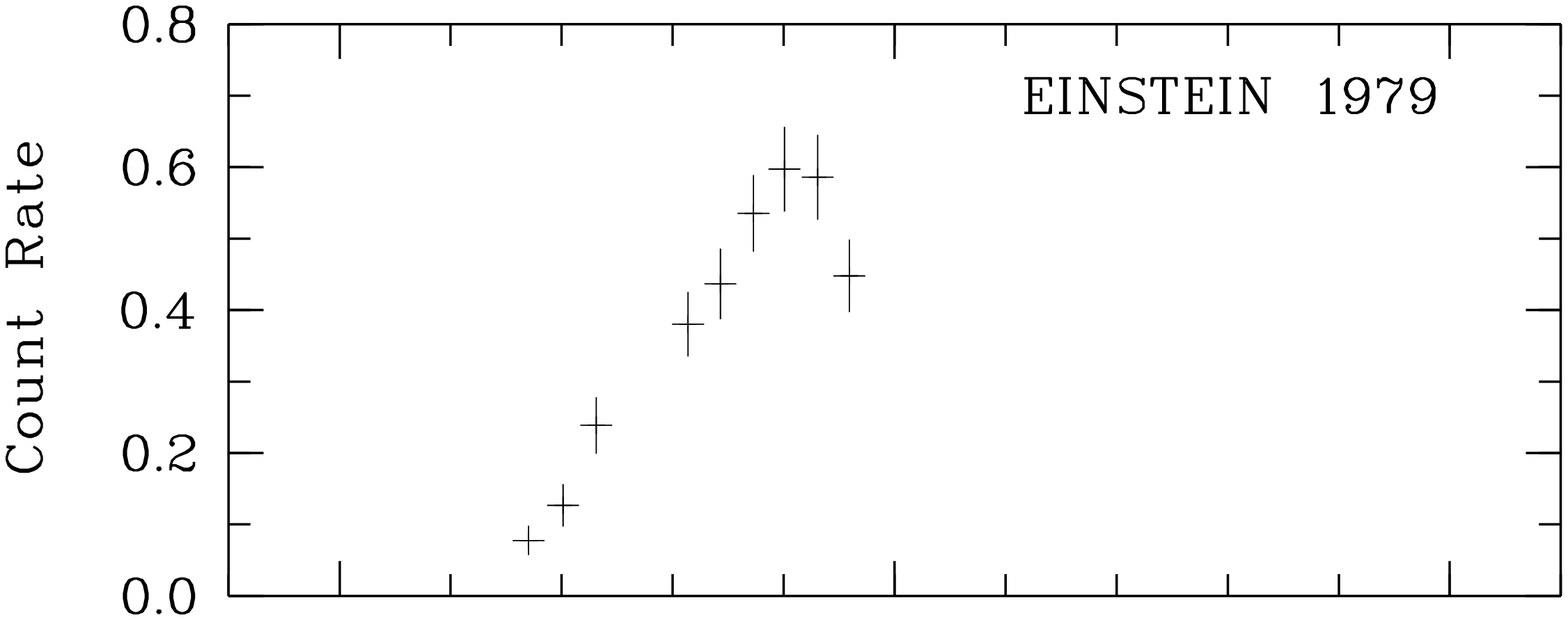}
\includegraphics[width=8.7cm]{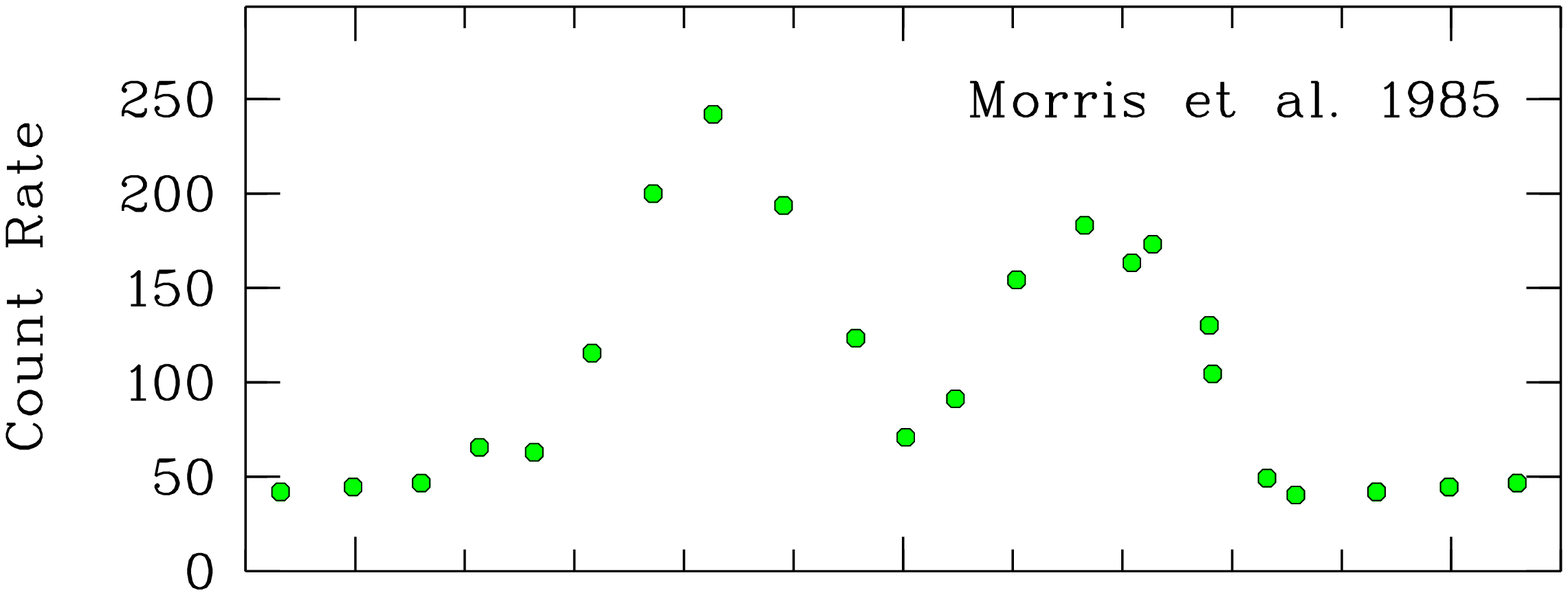}
\includegraphics[width=8.7cm]{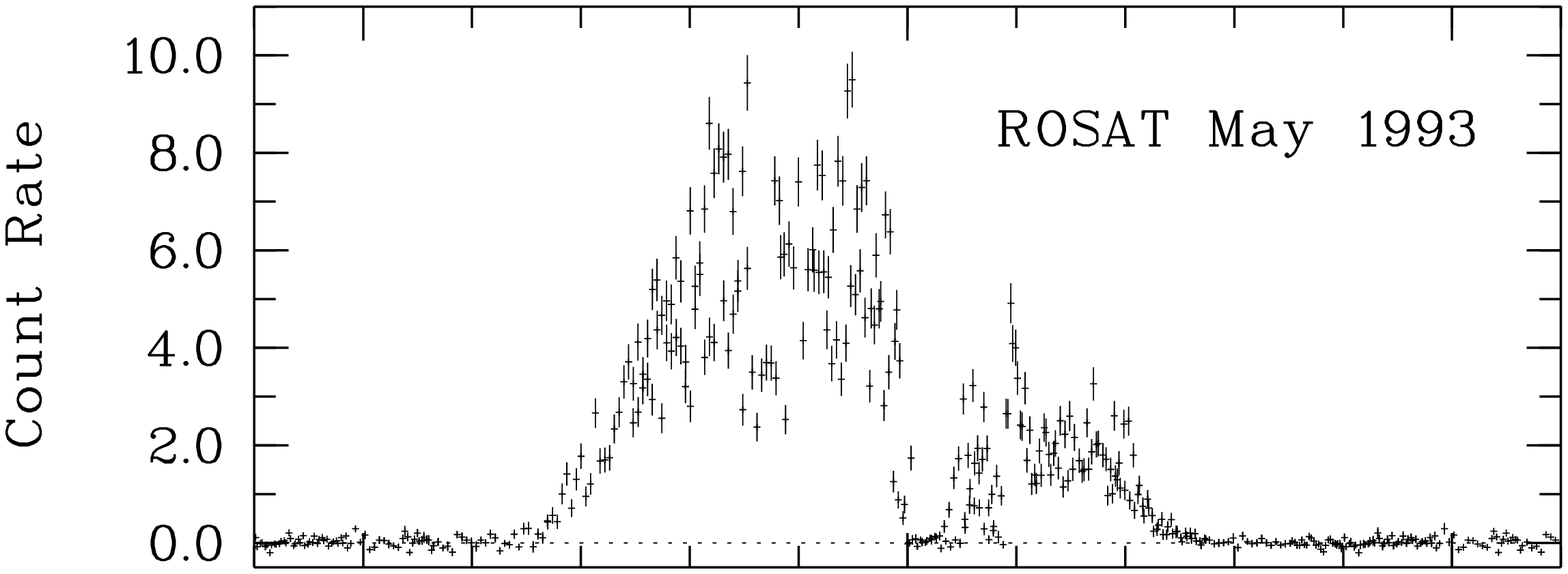}
\includegraphics[width=8.7cm]{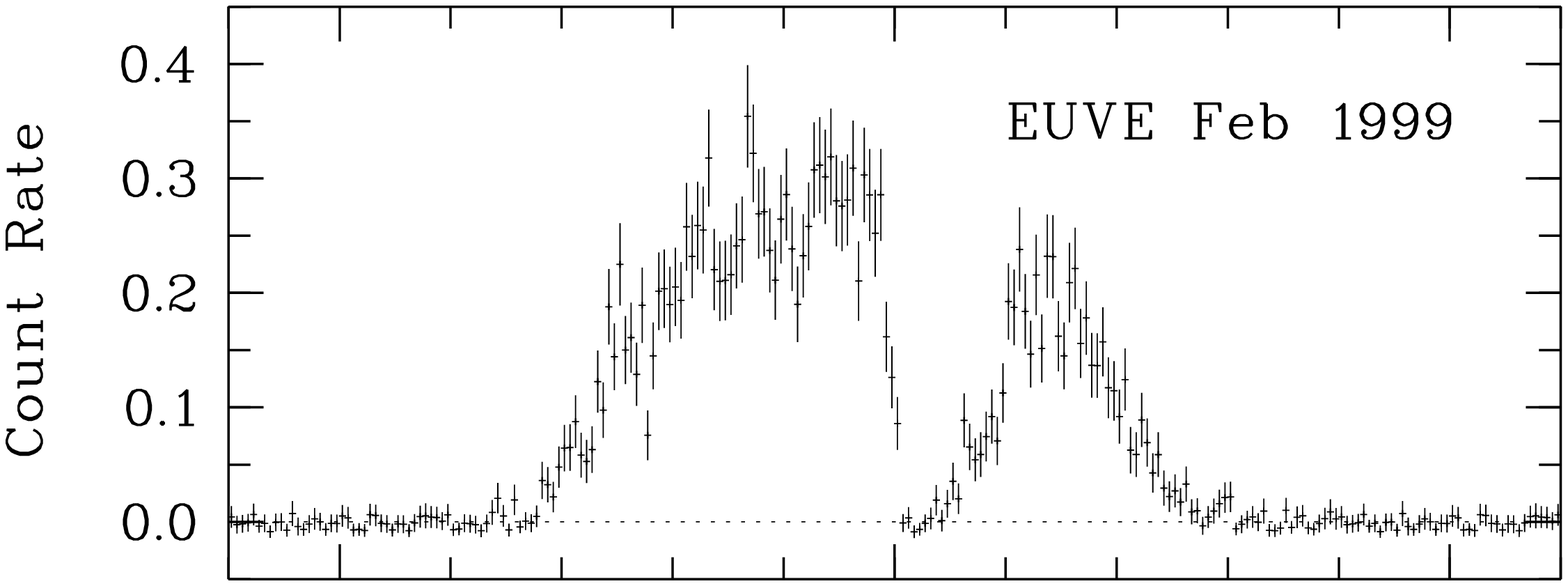}
\includegraphics[width=8.7cm]{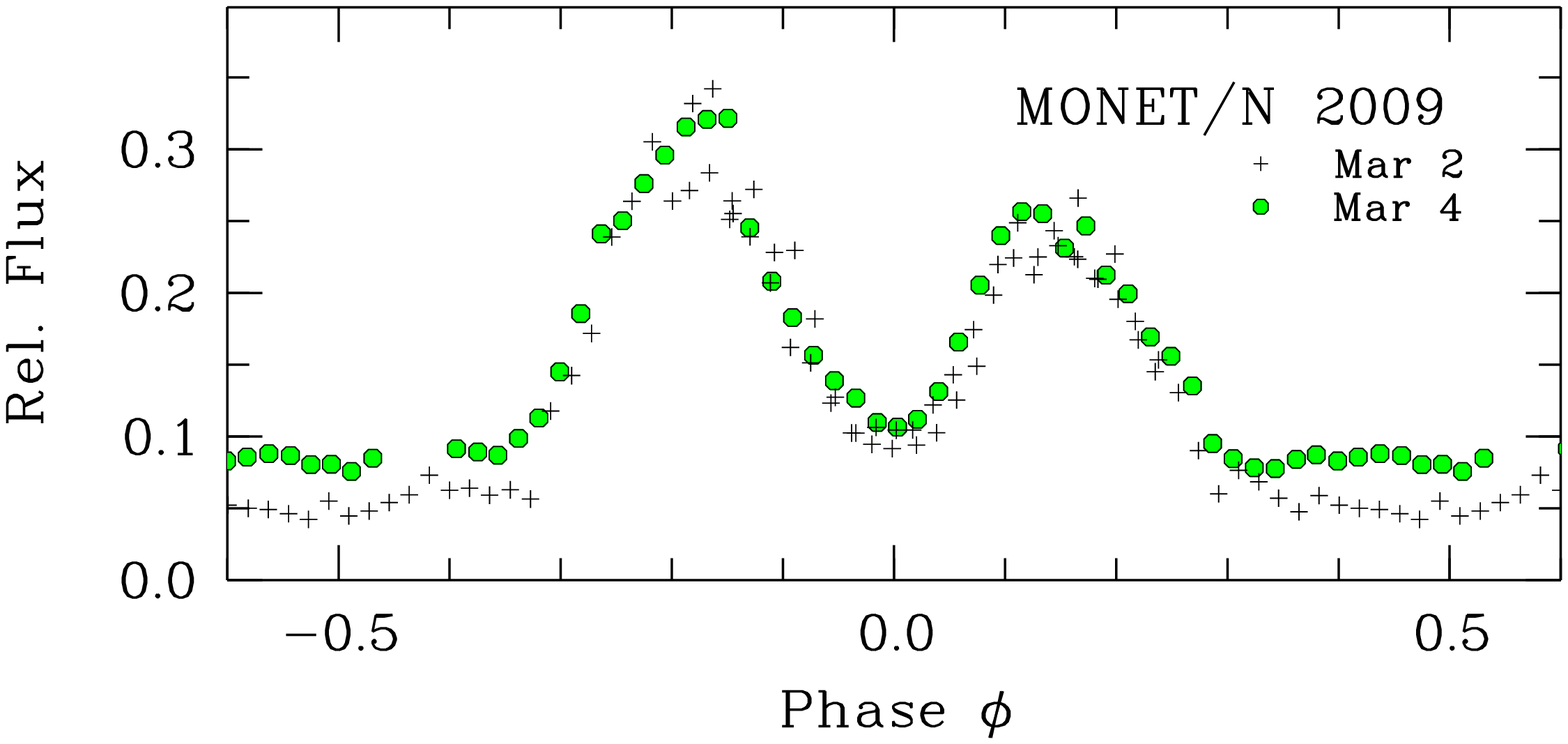}

\caption[ ]{Phased light curves based on the optical ephemeris of
  Eq.~(1). The Morris et al. (1987) and the MONET/N data were taken in
  white light. The ordinate for the latter is the flux of
  \ekuma\ relative to comparison star C1, and for all other panels, the
  ordinate is count rate in units of s$^{-1}$. The two MONET/N light
curves are for cycles $E=0$ and $E=24$.}
\label{fig:lc}
\end{figure}

EK~UMa was observed remotely with the MONET/N telescope of the
University of G\"ottingen at the McDonald Observatory in eight nights
between 26 February 2009 and 20 April 2009. All observations were
performed in white light, as was the 1985 discovery light curve of
Morris et al. (1987). The optical light curve possesses a bright phase
with two maxima that mark the times when one views the cyclotron
emission region at large angles to the magnetic field. They are
separated by a deep minimum that occurs when one views most directly
along the field. Our light curves, of which two examples are displayed
in Fig.~1 (bottom panel), are based on relative photometry with
respect to our comparison star C1, which is located 124 arcsec E and
178 arcsec N of EK~UMa and is about 2~mag
brighter\footnote{http://cas.sdss.org/dr7/en/tools/ provides the
SDSS magnitudes: EK\,UMa=SDSS J105135.14+540436.0,
  $g=18.38, r=18.99, i=19.04, z=19.19$; C1=SDSS J105149.27+540733.6,
  $g=18.58, r=17.11, i=16.30, z=15.86$; C2=SDSS J105138.99+540543.4,
  $g=20.69, r=19.22, i=18.43, z=17.96$.}.  During our runs,
\ekuma\ varied between $\sim18$ and $\sim20$\,mag, close to what was
estimated by \citet{morrisetal87}. A second comparison star C2, about
equal in brightness to the orbital average of EK~UMa and 34\,arcsec E
and 67\,arcsec N, was used to estimate the error in the photo\-metry
of EK~UMa. Following these authors, we used the deep central cyclotron
minimum as a fiducial mark for the ephemeris. 

The time of the minimum $T_\mathrm{min}$ and its error were determined
by two methods, (1)~by a parabola fit to the data points around
minimum, and (2)~graphically on plots of the individual light curves.
Method~1 has the advantage of yielding formal errors for the minimum
timings, but suffers from the varying asymmetry of the light
curves. In method~2, the center between descending and ascending parts
of the light curve is marked at different flux levels and
$T_\mathrm{min}$ determined by extrapolating any level-dependent shift
in the markings to the flux minimum. The timing error is estimated
from the scatter of the markings. This method provides insight into
the error sources and is the preferred one.  

Fifteen MONET/N timings of the central minimum are listed in Table~1,
along with the estimated errors. All timings are corrected to the
solar system barycenter, corrected for leap seconds, and quoted as
BJD(TT). Included in the table is the optical timing of
\citet{morrisetal87}, corrected for leap seconds. Its error of 0.003
days may represent an overestimate as judged from the light curve in
Fig.~3 of \citet{morrisetal87}. Also given in Table~1 are the $O-C$
values for the fits discussed below in days and in phase
units. Timings with weight $W=0$ are not included in the fits. The
band of the observation is quoted as `O' for optical and `X' for
X-ray/EUV data. Note that there is no uncertainty in the cycle numbers
given in the first column.

The shapes of our 2009 light curve and the 1985 light curve of Morris
et al. are very similar (Fig.~\ref{fig:lc}, bottom panel and second
from top). Some night-to-night variability is present and is
illustrated by the bright and faint phase intervals
varying independently in flux and subtle changes also occurring in the
shapes of the light curves (Fig.~1, bottom panel). The bright phase
extends over about 65\% of the orbital period, indicating that the
main accreting pole is located in the upper hemisphere of the white
dwarf. The bright-phase parameters are given in Table~2. The existence
of the X-ray absorption dip requires that the inclination exceeds the
co-latitude of the accretion spot.

\begin{figure*}[t]
\includegraphics[width=10.5cm]{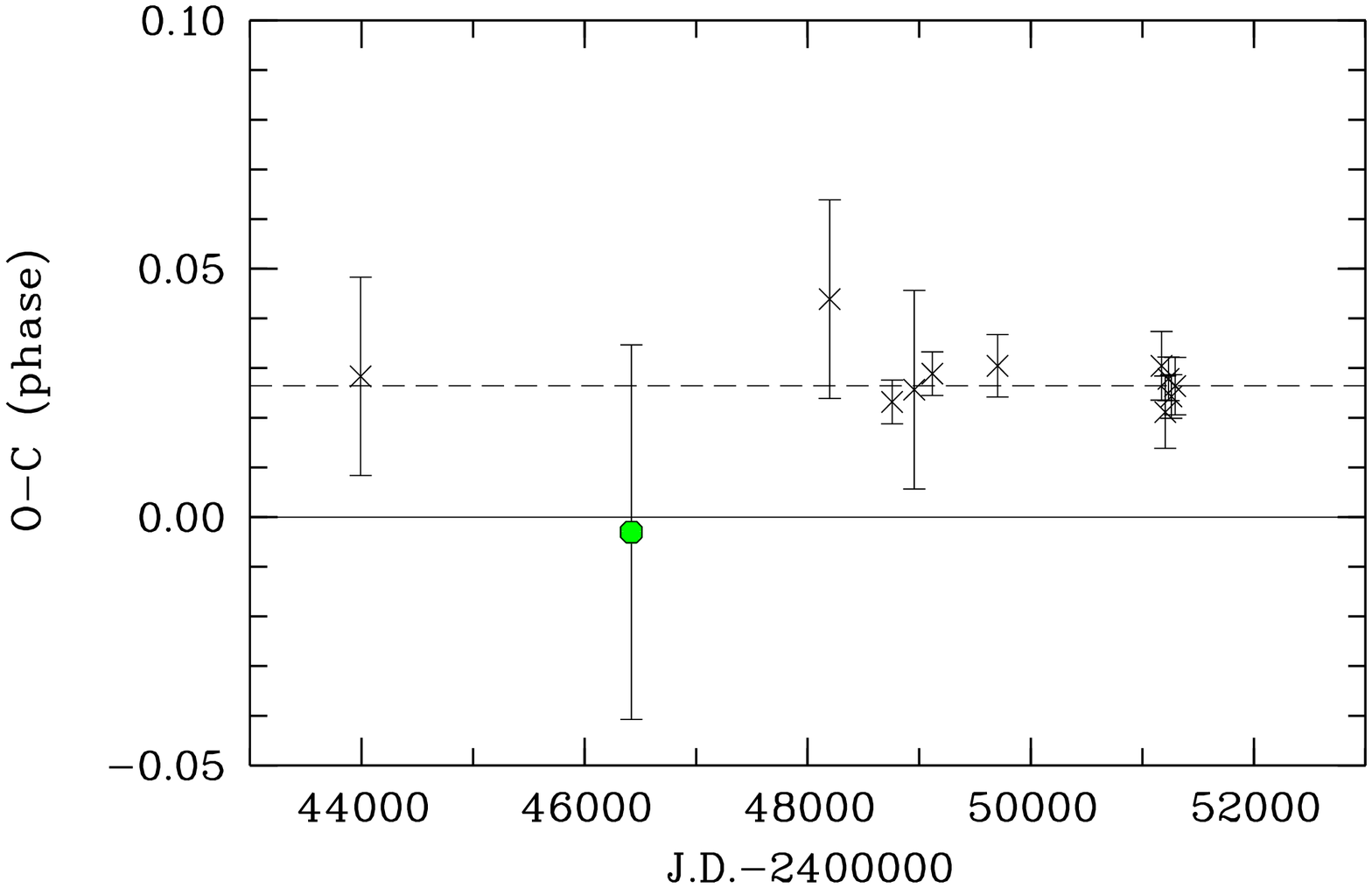}
\hspace{1mm}
\includegraphics[width=7.25cm]{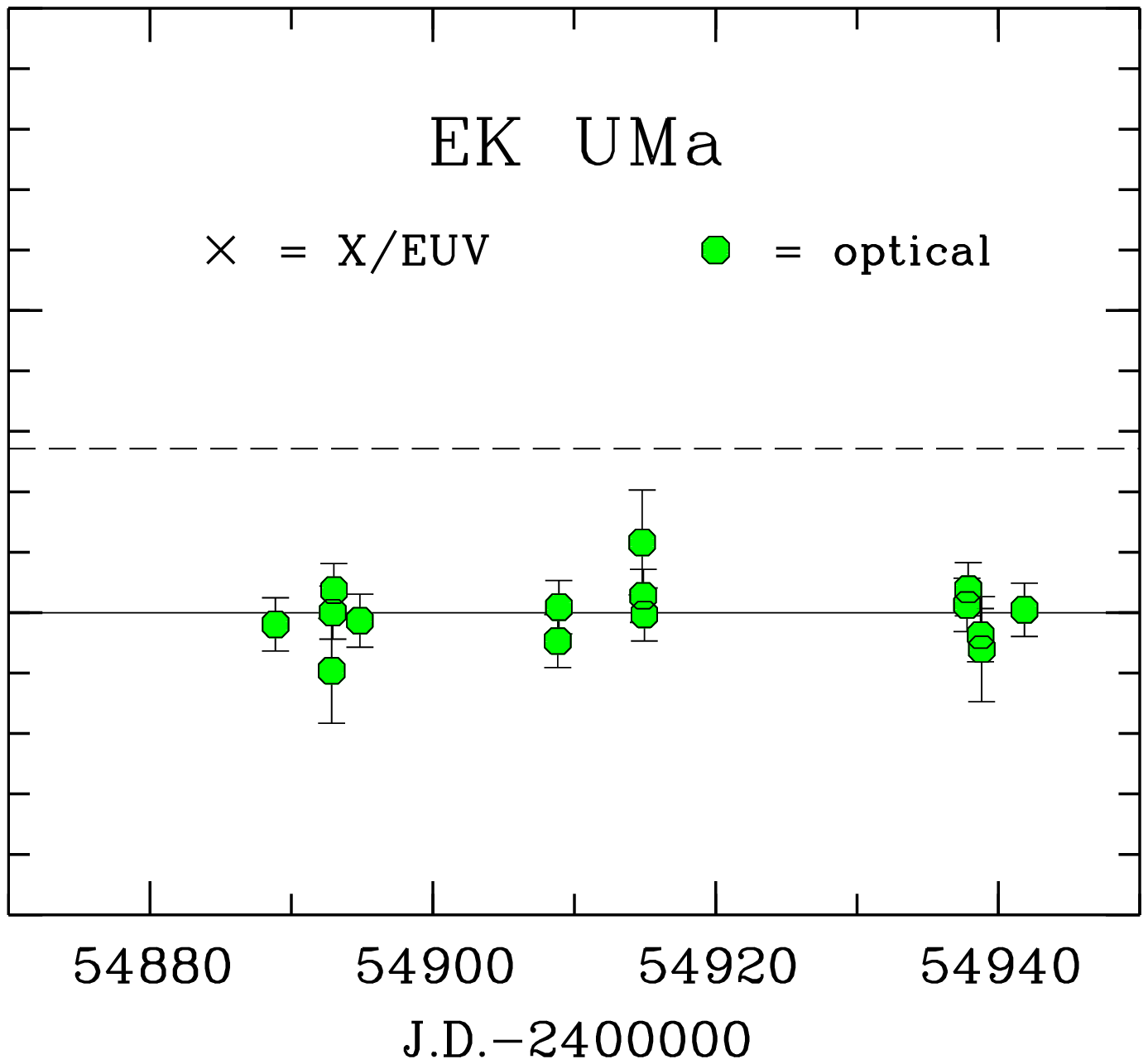}
\caption[ ]{$O-C$ diagrams for timings of the optical cyclotron minima
  and the X-ray absorption dips before 2000 (left) and in 2009
  (right). The solid and dashed lines represent the linear ephemeris of
  Eq.~(1) with $\Delta T_{\rm o}=0$ and $\Delta T_{\rm x}/P=0.0264$,
  respectively. The scale is changed between the two panels.}
\label{fig:drm}
\vspace*{5mm}
\end{figure*}

\begin{table*}[t]
\caption{Timings of optical minima and X-ray/EUV dips of EK~UMa quoted
  as terrestrial time at the solar-system barycenter. }
\label{tab:diplist}
\begin{tabular}{rccccrrrrcl} 
\hline \hline \noalign{\smallskip}
Cycle     & BJD(TT)        & \multicolumn{2}{c}{Error}  & W    & \multicolumn{2}{c}{$O-C$} & \multicolumn{2}{c}{$O-C-\Delta T_{\rm o,x}$}  & B & Instrument \\
Number     & (days)         & (days)      & (phase)     &      & (days)      & (phase)     &   (days)    &  (phase)    &   &      \\
\noalign{\smallskip} \hline
\noalign{\smallskip}
$-137044 $&$ 43991.898920 $&$  0.001590 $&$  0.019989 $&$  0 $&$  0.002254 $&$  0.028331 $&$  0.000151 $&$  0.001894 $& X & {\it EINSTEIN}$^{~(1)}$\\  
$-106532 $&$ 46418.943640 $&$  0.003000 $&$  0.037715 $&$  1 $&$ -0.000242 $&$ -0.003038 $&$ -0.000242 $&$ -0.003038 $& O & Morris et al.\\           
$ -84181 $&$ 48196.835820 $&$  0.001590 $&$  0.019989 $&$  0 $&$  0.003491 $&$  0.043886 $&$  0.001388 $&$  0.017449 $& X & {\it RASS}$^{~(2)}$\\      
$ -77149 $&$ 48756.187740 $&$  0.000350 $&$  0.004400 $&$  1 $&$  0.001844 $&$  0.023188 $&$ -0.000258 $&$ -0.003249 $& X & {\it ROSAT}\\             
$ -74652 $&$ 48954.809360 $&$  0.001590 $&$  0.019989 $&$  0 $&$  0.002040 $&$  0.025649 $&$ -0.000063 $&$ -0.000788 $& X & {\it ROSAT}$^{~(1)}$ \\    
$ -72610 $&$ 49117.238510 $&$  0.000350 $&$  0.004400 $&$  1 $&$  0.002296 $&$  0.028867 $&$  0.000193 $&$  0.002430 $& X & {\it ROSAT}\\             
$ -65258 $&$ 49702.046290 $&$  0.000500 $&$  0.006286 $&$  1 $&$  0.002423 $&$  0.030456 $&$  0.000320 $&$  0.004019 $& X & {\it EUVE}\\              
$ -46777 $&$ 51172.099370 $&$  0.000550 $&$  0.006914 $&$  1 $&$  0.002422 $&$  0.030453 $&$  0.000319 $&$  0.004016 $& X & {\it EUVE}\\              
$ -46359 $&$ 51205.348030 $&$  0.000580 $&$  0.007292 $&$  1 $&$  0.001681 $&$  0.021132 $&$ -0.000422 $&$ -0.005305 $& X & {\it EUVE}\\              
$ -45992 $&$ 51234.541220 $&$  0.000350 $&$  0.004400 $&$  1 $&$  0.002215 $&$  0.027842 $&$  0.000112 $&$  0.001405 $& X & {\it EUVE}\\              
$ -45673 $&$ 51259.915480 $&$  0.000350 $&$  0.004400 $&$  1 $&$  0.001931 $&$  0.024282 $&$ -0.000171 $&$ -0.002155 $& X & {\it EUVE}\\              
$ -45247 $&$ 51293.801400 $&$  0.000460 $&$  0.005783 $&$  1 $&$  0.002098 $&$  0.026374 $&$ -0.000005 $&$ -0.000063 $& X & {\it EUVE}\\              
$    -51 $&$ 54888.870790 $&$  0.000350 $&$  0.004400 $&$  1 $&$ -0.000154 $&$ -0.001938 $&$ -0.000154 $&$ -0.001938 $& O & {\it MONET/N}\\           
$     -1 $&$ 54892.847380 $&$  0.000690 $&$  0.008674 $&$  1 $&$ -0.000765 $&$ -0.009621 $&$ -0.000765 $&$ -0.009621 $& O & {\it MONET/N}\\           
$      0 $&$ 54892.927690 $&$  0.000350 $&$  0.004400 $&$  1 $&$  0.000001 $&$  0.000009 $&$  0.000001 $&$  0.000009 $& O & {\it MONET/N}\\           
$      1 $&$ 54893.007530 $&$  0.000350 $&$  0.004400 $&$  1 $&$  0.000297 $&$  0.003730 $&$  0.000297 $&$  0.003730 $& O & {\it MONET/N}\\           
$     24 $&$ 54894.836640 $&$  0.000350 $&$  0.004400 $&$  1 $&$ -0.000106 $&$ -0.001331 $&$ -0.000106 $&$ -0.001331 $& O & {\it MONET/N}\\           
$    200 $&$ 54908.836120 $&$  0.000350 $&$  0.004400 $&$  1 $&$ -0.000374 $&$ -0.004699 $&$ -0.000374 $&$ -0.004699 $& O & {\it MONET/N}\\           
$    201 $&$ 54908.916110 $&$  0.000350 $&$  0.004400 $&$  1 $&$  0.000072 $&$  0.000907 $&$  0.000072 $&$  0.000907 $& O & {\it MONET/N}\\           
$    275 $&$ 54914.803220 $&$  0.000690 $&$  0.008674 $&$  1 $&$  0.000925 $&$  0.011623 $&$  0.000925 $&$  0.011623 $& O & {\it MONET/N}\\           
$    276 $&$ 54914.882060 $&$  0.000350 $&$  0.004400 $&$  1 $&$  0.000220 $&$  0.002772 $&$  0.000220 $&$  0.002772 $& O & {\it MONET/N}\\           
$    277 $&$ 54914.961360 $&$  0.000350 $&$  0.004400 $&$  1 $&$ -0.000024 $&$ -0.000296 $&$ -0.000024 $&$ -0.000296 $& O & {\it MONET/N}\\           
$    564 $&$ 54937.790620 $&$  0.000350 $&$  0.004400 $&$  1 $&$  0.000102 $&$  0.001282 $&$  0.000102 $&$  0.001282 $& O & {\it MONET/N}\\           
$    565 $&$ 54937.870370 $&$  0.000350 $&$  0.004400 $&$  1 $&$  0.000308 $&$  0.003872 $&$  0.000308 $&$  0.003872 $& O & {\it MONET/N}\\           
$    576 $&$ 54938.744750 $&$  0.000350 $&$  0.004400 $&$  1 $&$ -0.000296 $&$ -0.003725 $&$ -0.000296 $&$ -0.003725 $& O & {\it MONET/N}\\           
$    577 $&$ 54938.824110 $&$  0.000690 $&$  0.008674 $&$  1 $&$ -0.000480 $&$ -0.006038 $&$ -0.000480 $&$ -0.006038 $& O & {\it MONET/N}\\           
$    615 $&$ 54941.847300 $&$  0.000350 $&$  0.004400 $&$  1 $&$  0.000037 $&$  0.000463 $&$  0.000037 $&$  0.000463 $& O & {\it MONET/N}\\           
\noalign{\smallskip} \hline 
\end{tabular}

$^{(1)}$ Based on the start of the bright phase at $\phi=-0.37\pm 0.02$ relative to the center of the X-ray dip. Not used in deriving the ephemeris. \\
$^{(2)}$ \rosat\ All Sky Survey. Not used in deriving the ephemeris. 
\end{table*}

\begin{table}[b]
\caption{Phases $\phi$ for the the central minima and the lengths of
  the bright intervals, with $\phi=0$ referring to the optical
  minimum. }
\label{tab:diplist}
\begin{tabular}{l@{\hspace{10mm}}c@{\hspace{8mm}}c@{\hspace{6mm}}c@{\hspace{6mm}}l} 
\hline \hline \noalign{\smallskip}
Observation & $\phi_{\rm min/dip}$ & \multicolumn{3}{c}{Bright phase interval~$^{(1)}$} \\
 &  &  $\phi_{\rm start}$ &  $\phi_{\rm end}$ &  $\phi_{\rm center}$\\
\noalign{\smallskip} \hline
\noalign{\smallskip}
Optical 1987
            & $0.000$ & $-0.34$ & $0.32$ & $-0.01$ \\ 
{\it ROSAT} 1992/1993 & $0.026$ & $-0.34$ & $0.26$ & $-0.04$\\ 
{\it EUVE} 1994/1999  & $0.026$ & $-0.33$ & $0.27$ & $-0.03$\\ 
{\it MONET/N} 2009 & $0.000$ & $-0.33$ & $0.31$ & $-0.01$\\ 
\noalign{\smallskip} \hline 
\end{tabular}

$^{(1)}$ Errors typically 0.01, except for the
1987 data, where they are 0.02.
\end{table}

\section{X-ray and EUV light curves}

We obtained soft X-ray light curves from archival data taken with the
{\it ROSAT} satellite in May 1992, Nov 1992, and May 1993, and extreme
ultraviolet light curves from archival data taken with the {\it EUVE}
deep survey (DS/S) instrument in Dec 1994, Dec 1998, and Jan, Feb,
Mar, and Apr 1999. The \einstein\ soft X-ray light curve of 1979 was
taken from Fig.~2 of \citet{morrisetal87}. We show the
\einstein\ light curve and phase-folded examples of the \rosat\ and
\euve\ light curves in Fig.~1. The individual light curves of the
three \rosat\ and the six \euve\ observation periods are very similar
in shape, with some differences in phase coverage and brightness
levels. The May 1993 \rosat\ light curve has the best phase coverage
of the set of three, and the Feb 1999 \euve\ light curve has the
highest count rate of the set of six.  The X-ray bright phase
intervals coincide approximately with the optical ones and are
interrupted by a dip, which occurs when the line of sight crosses the
accretion stream at some distance from the white dwarf. We note that
the X-ray dips in EK~UMa vary in length and possess an ingress that is
better defined than the egress. This can be understood in terms of an
accretion stream that forms an extended curtain, in which some matter
attaches to the field earlier than the bulk of the material and
absorbs X-rays after the line of sight has passed the proper accretion
stream. We disregard soft X-ray minima with obvious post-dip
absorption, a caveat that is particularly important for the {\it
  EUVE\/} DS/S light curves with their low count rates. The mean
timing for each data set (month) of the \rosat\ and the \euve\ DS/S
observations is given in Table~1 along with the estimated timing
error. The {\it EINSTEIN\/} data and the Nov 1992 {\it ROSAT\/} data
do not cover the X-ray absorption dip, and the corresponding dip
timings are estimated from the well-defined start of the bright
phase. The last two and the timing from the ROSAT All-Sky-Survey
(RASS) are not included in the fit.

\section{Results}

The fifteen new MONET/N timings of 2009 combined with the 1985 timing
of \citet{morrisetal87} define an alias-free ephemeris with an optical
period $P_{\rm opt}=0.079544025(28)$ days, where the number in
brackets refers to the uncertainty in the last digits. This
uncertainty arises almost entirely from the large error assigned to
the 1985 timing by \citet{morrisetal87}. Fitting the eight timings
from the pointed {\it ROSAT} and {\it EUVE} data yields $P_{\rm
  x}=0.079544022(3)$ days. Both periods agree within the uncertainties
with the pre-1995 period of Schwope et al. (1995). There is a
significant difference, however, in the epochs of the optical and the
X-ray/EUV ephemerides, indicating that the X-ray/EUV dips occur about
3~min later than the optical minima.  We proceed with a
three-parameter fit to the combined set of 24 optical and X-ray/EUV
timings, $T_{\rm min}= T_0+\Delta T_{\rm o,x}+P\,E$.  The three
parameters are the common period $P$, the epoch $T_0$, and an offset
$\Delta T_{\rm x}$, which we subtract from the X-ray/EUV dip times
prior to the common fit. Our reference is the optical ephemeris and
the corresponding quantity $\Delta T_{\rm o}$ for the optical timings
is set to zero.  The resulting linear ephemeris, applicable to the
optical \textit{and} the X-ray data, is
\begin{equation}
T_{\rm min}=2454892.92769(8)+\Delta T_{\rm o,x}+0.0795440225(24)\,E,
\end{equation}
with $T_{\rm min}$ given as barycentrically corrected terrestrial
(ephemeris) times, $\Delta T_{\rm o}\equiv 0$, $\Delta T_{\rm
  x}=0.00210(14)$ days, and $\chi^2$ of 10.2 for 21 degrees of
freedom.
An added quadratic term $CE^2$ in the ephemeris turns out to be
insignificant with $C=-(1.1\pm3.7)\,10^{-14}$\,days and $\dot
P=-(2.7\pm9.3)\,10^{-13}$\,s\,s$^{-1}$. The error in $C$ limits the period
variation over the well-observed time interval from 1992 to 2009 with
$\Delta E=77759$ cycles to $\Delta P = 2C\Delta E=-(0.14\pm0.50)$\,ms.

The offset between the cyclotron minima and the X-ray
absorption dips indicates that the magnetic funnel near the white
dwarf leads the X-ray absorption region by
$9.5^\circ\pm0.7^\circ$. The sense of curvature of the stream is as
expected if the accreting pole precedes the coupling region in the
magnetosphere.

The measured period is the rotation period of the white dwarf, provided
the accretion spot does not wander over the white dwarf. The mean
start and stop phases of the bright part of the light curve appear
constant over 24 years (Fig.~1 and Table~2), and the ensuing lack of a
change in the accretion geometry suggests that what we measure is the
orbital period and that the system is tightly synchronized. A final
decision on the degree of synchronization requires long-term
spectroscopic observations of emission or absorption lines from the
secondary star.

\section{Discussion}

We find that the optical and X-ray data of \ekuma\ define an ephemeris
that is linear over at least 17 years. A quadratic term is
insignificant and limits the period variation over this time span to
$-1.1<\Delta P<0.9$\,ms (2-$\sigma$ limit). With long-term coverage
becoming available for an increasing number of CVs, measuring period
variations in the ms or sub-ms range becomes feasible. In the
eclipsing pre-CV NN~Ser, \citet{brinkworthetal06} find $\dot
P=-(9.06\pm0.06)\,10^{-12}$\,s\,s$^{-1}$ over 45\,000 cycles and a
still higher $\dot P=-(2.85\pm0.15)\,10^{-11}$\,s\,s$^{-1}$ over some
3\,000 cycles. The eclipsing polars DP~Leo
\citep{pandeletal02,schwopeetal02} and HU~Aqr \citep{schwarzetal09}
showed decreases in the orbital periods with $\dot
P\simeq-5\,10^{-12}$\,s\,s$^{-1}$ over 120\,000 cycles and $\dot
P=-(7.3\pm0.5)\,10^{-13}$\,s\,s$^{-1}$ over 60\,000 cycles, with
superposed possibly sinusoidal variations in the latter. A sinusoidal
modulation in the $O-C$ values has previously been seen in the
eclipsing dwarf nova U~Gem \citep[e.g.][]{beuermannpakull84}. While
star spot cycles may account for the quasi-periodic variations, there
is no entirely convincing explanation of the secular period changes
emphasizing the need for further studies meant to differentiate
between typical behavior and idiosyncrasies. We have demonstrated that
subtle period variations -- or their absence -- are easily detected
using a small, remotely operated telescope.

\begin{acknowledgements} The MONET Monitoring Network of 1.2-m
optical telescopes is funded by the Alfried Krupp von Bohlen und
Halbach Foundation and provides part of the observation time to
astronomical projects at high schools. This research has made use of
archival data obtained with the ROSAT X-ray satellite and the EUVE
extreme ultraviolet satellite. The data were retrieved from the ROSAT
Archive at the MPE at Garching/Munich and the High Energy Astrophysics
Science Archive Research Center (HEASARC) at NASA's Goddard Space
Flight Center.
\end{acknowledgements}

\bibliographystyle{aa}

\begin{thebibliography}{29}
\expandafter\ifx\csname natexlab\endcsname\relax\def\natexlab#1{#1}\fi

\bibitem[Beuermann \& Pakull(1984)]{beuermannpakull84} Beuermann,
  K., \& Pakull, M.~W.\ 1984, \aap, 136, 250
\bibitem[Brinkworth et al.(2006)]{brinkworthetal06} Brinkworth, C.~S., 
Marsh, T.~R., Dhillon, V.~S., \& Knigge, C.\ 2006, \mnras, 365, 287 
\bibitem[Morris et al.(1987)]{morrisetal87} Morris, S.~L., Schmidt, 
G.~D., Liebert, J. et al.\ 1987, \apj, 314, 641 
\bibitem[Clayton \& Osborne(1994)]{claytonosborne94} Clayton K.~L., \&
  Osborne, J.~P.\ 1994, \mnras, 268, 229
\bibitem[Cropper et al.(1990)]{cropperetal90} Cropper, M., Mason,
  K.~O., \& Mukai, K.\ 1990, \mnras, 243, 565
\bibitem[Pandel et al.(2002)]{pandeletal02} Pandel, D., Cordova, 
F.~A., Shirey, R.~E. et al.\ 2002, \mnras, 332, 116 
\bibitem[Schwarz et al.(2009)]{schwarzetal09} Schwarz, R., Schwope,
  A.~D., Vogel, J. et al.\ 2009, \aap, 496, 833
\bibitem[Schwope et al.(1995)]{schwopeetal95} Schwope, A.~D.,
  Beuermann, K., Burwitz, V., Mantel, K.-H., \& Schwarz, R.\ 1995,
  Cataclysmic Variables, ASSL, 205, 389
\bibitem[Schwope et al.(2002)]{schwopeetal02} Schwope, A.~D.,
  Hambaryan, V., Schwarz, R., Kanbach, G., G\"ansicke,
  B.~T.\ 2002, \aap, 392, 541
\end{thebibliography}

\end{document}